\documentclass[published]{JHEP3}

\JHEP{04(2004)056}
\received{March 31, 2004}
\accepted{April 22, 2004\\ \quad\\
{\rm {\bf JHEP} 04(2004)056,}\\
{\rm hep-th/0403187,}\\
{\rm HUTP-04/A012,}\\ {\rm IFT-P.007/2004,}\\ {\rm HEP-UK-0020}}


\usepackage{epsfig}

\skip\footins = 1\bigskipamount plus 2pt minus 4pt

\def\IR{\mathbb{R}}
\def\N{{\cal N}}

\def\tilde{\widetilde}
\def\eqref#1{(\ref{#1})}
\def\ket#1{\left\vert #1\right\rangle}
\def\eqn#1#2{\begin{equation}#2\label{#1}\end{equation}}
\def\Tr{\mathop{\rm Tr}\nolimits}
\def\a{\alpha}

\def\s{\sigma}
\def\l{\lambda}
\def\g{\gamma}
\def\d{\delta}
\def\de{{\rm d}}
\def\e{\epsilon}
\def\ad{{\dot a}}
\def\p{\partial }

\newcommand{\U}{\mathop{\rm {}U}\nolimits}
\newcommand{\SL}{\mathop{\rm SL}\nolimits}
\newcommand{\GL}{\mathop{\rm GL}\nolimits}
\newcommand{\PSU}{\mathop{\rm PSU}\nolimits}

\title{Cubic twistorial string field theory}

\author{Nathan Berkovits\\
  Instituto de F\'{\i}sica Te\'{o}rica, Universidade Estadual Paulista\\
  Rua Pamplona 145, 01405-900, S\~ao Paulo, SP, Brasil\\
  E-mail: \email{nberkovi@ift.unesp.br}}

\author{Lubo\v{s} Motl\\
  Jefferson Physical Laboratory\\
  Harvard University, Cambridge, MA 02138, U.S.A.\\
  E-mail: \email{motl@feynman.harvard.edu}}

\abstract{Witten has recently proposed a string theory in twistor
  space whose D-instanton contributions are conjectured to compute
  ${\cal N}=4$ super-Yang-Mills scattering amplitudes. An alternative
  string theory in twistor space was then proposed whose open string
  tree amplitudes reproduce the D-instanton computations of maximal
  degree in Witten's model.

In this paper, a cubic open string field theory action is constructed
for this alternative string in twistor space, and is shown to be
invariant under parity transformations which exchange MHV and googly
amplitudes. Since the string field theory action is gauge-invariant
and reproduces the correct cubic super-Yang-Mills interactions, it
provides strong support for the conjecture that the string theory
correctly computes $N$-point super-Yang-Mills tree amplitudes.}

\begin{document}

\section{Introduction}

In a recent paper~\cite{wt}, Witten showed that the simple holomorphic
form of maximal helicity-violating (MHV) tree amplitudes in $\N=4$
super-Yang-Mills theory~\cite{mhvone,mhvtwo,mhvthree} can be
generalized to tree amplitudes with less than maximal
helicity-violation. He showed that just as MHV tree amplitudes (i.e.\
amplitudes with two negative-helicity gluons) are described by curves
in twistor space of degree one, tree amplitudes with $d+1$
negative-helicity gluons are described by curves in twistor space of
degree $d$.

Witten then proposed a string theory based on twistor worldsheet
variables whose D-instanton contributions were used to compute $\N=4$
super-Yang-Mills scattering amplitudes. In this string theory,
Yang-Mills tree amplitudes with $d+1$ negative-helicity gluons were
computed by D-instantons of degree less than or equal to $d$. Since
the Yang-Mills tree amplitudes are not described by perturbative
amplitudes in the string theory, it is not straightforward to check
properties such as factorization and unitarity. This makes it
complicated to prove the conjecture that the D-instanton contributions
correctly reproduce the super-Yang-Mills tree amplitudes.

Last month, an alternative string theory based on twistor worldsheet
variables was constructed~\cite{berkovits} in which $\N=4$
super-Yang-Mills tree amplitudes were conjectured to be computed by
tree amplitudes in the string theory.  Assuming that only D-instantons
of maximal degree contribute in the model of Witten, these two string
theories give the same prescription for super-Yang-Mills tree
amplitudes. Note that for MHV amplitudes, it is clear that only
D-instantons of maximal degree contribute in Witten's model. And it
was recently shown for ``googly'' amplitudes containing two
positive-helicity gluons that only maximal degree D-instantons are
needed~\cite{spradlin,roiban}.\footnote{Shortly after our paper was
posted on the bulletin board, there appeared a paper~\cite{roibantwo}
proving on-shell parity symmetry and showing that six-point amplitudes
with three negative-helicity and three positive-helicity gluons also
only require maximal degree D-instantons.}

In this paper, we construct a cubic open string field theory action
for the alternative string theory in twistor
space~\cite{berkovits}. The cubic action is gauge-invariant and
reproduces the usual cubic supersymmetric Yang-Mills interaction
terms. Furthermore, it is invariant under parity transformations which
exchange MHV and googly amplitudes\footnote{It was recently shown by
  Edward Witten in independent work that twistor calculations are
  invariant under parity transformations~\cite{wittenp}. There has
  also been some recent work~\cite{agavafa} by Aganagic and Vafa who
  explain parity symmetry in the context of the model
  of~\cite{nejvafa}.}.  Gauge-fixing is straightforward using the
$b_0=0$ Siegel gauge, so the action can be used to define string
Feynman diagrams and compute $N$-point tree amplitudes. Using the
standard open string field theory result that the cubic vertex
correctly covers moduli space~\cite{gm, gmw, zwiebach}, these field
theory computations agree with the first-quantized worldsheet theory
computations. Since the string Feynman diagrams are expected to
be factorizable with the appropriate poles, and since the cubic
Yang-Mills amplitudes are correctly reproduced, the field theory
action constructed here gives strong evidence for the conjecture that
the open string amplitudes correctly compute the $N$-point
super-Yang-Mills tree amplitudes.

In section~\ref{reviewed} of this paper, the alternative string theory
in twistor space is reviewed. In section~\ref{csft}, a cubic
twistorial string field theory action is constructed. In
section~\ref{paritysection}, the action is proven to be invariant
under parity transformations. And in section~\ref{outlook}, we discuss
conclusions and possible applications.

\section{Review of the open twistorial string}\label{reviewed}

\subsection{Worldsheet action}

In the open twistorial string theory proposed in~\cite{berkovits}, the
left-moving worldsheet variables consist of the real twistor variables
$Z_L^I = (\l_L^a,\mu_L^\ad,\psi_L^A)$ for $a,\ad=1$ to 2 and $A=1$ to
4, their conjugate momenta $Y_{L I}$, and a left-moving current
algebra $j_L^k$ for $k=1$ to $\dim(G)$.  Before twisting, the $Z_L^I$
variables have conformal weight zero and the $Y_{L I}$ variables have
conformal weight one.  One also has the right-moving variables
$Z_R^I$, $Y_{R I}$ and $j_R^k$, which satisfy the open string boundary
conditions
\eqn{boundary}
{Z_L^I = Z_R^I\,,\qquad 
Y_{LI}= Y_{RI}\,,\qquad
j_L^k = j_R^k.}
\pagebreak[3]Although the formalism resembles heterotic string
constructions~\cite{nair} because of the current algebra, it seems
necessary to work with open strings because the OPE of the vertex
operators should lead to the factors $1/(z_i-z_{i+1})$ whose dimension
is one, while the total dimension of the closed string vertex
operators is two.

Since $Z^I$ and $Y_I$ are real twistor variables, the target space on
which the super-Yang-Mills theory is defined has signature $(2,2)$ and
the worldsheet has Minkowski signature $(1,1)$.  Although one would
prefer to have a string theory defined in Minkowski signature $(3,1)$,
one can use this string theory to compute super-Yang-Mills scattering
amplitudes in signature $(2,2)$, and then Wick-rotate the results to
Minkowski signature $(3,1)$.  For the case of a $\U(M)$ gauge group,
the currents $j^k$ may be written as composite fermionic bilinear
operators $\alpha^i \beta_j$ for $i,j=1\dots M$, analogous to the 1-5
strings of~\cite{wt}. But at least at the classical level, the
formalism can be generalized to an arbitrary gauge group represented
by a current algebra.

The worldsheet action for the matter fields is
\eqn{action}
{S=\int \de^2 z \left[ Y_{L I} \nabla_{\bar z} Z_L^I +
Y_{R I} \nabla_z Z_R^I\right] + S_C}
where $S_C$ is the action for the left and right-moving current
algebras, and the covariant derivatives
\eqn{covderiv}
{\nabla_z = \p_z - A_z\,,\qquad
\nabla_{\bar z} = \p_{\bar z} - A_{\bar z}}
include a $\GL(1,\IR)\equiv\GL(1)$ worldsheet gauge field $A_\mu$
which is defined such that $Y_I$ has $-1$ $\GL(1)$ charge and $Z^I$ has
$+1$ $\GL(1)$ charge.

\subsection{Physical states}

Using the Virasoro and $\GL(1)$ generators together with their ghosts
$(b,c)$ of conformal weight $(2,-1)$ and $(u,v)$ of conformal weight
$(1,0)$, one can construct the BRST operator
\eqn{brst}
{Q=\int \de z \, [ cT + v J + c u \p v +  c b \p c]}
where the matter contribution to the stress tensor and $\GL(1)$ current is
\eqn{stress}
{T= Y_I \p Z^I + T_C\,, \qquad J= Y_I Z^I\,,}
and $T_C$ is the stress tensor for the current algebra. Note that
$Q^2=0$ when $T_C$ has central charge $+28$, which cancels $c=-26$ of
the $(b,c)$ system and $c=-2$ of the $(u,v)$ system. (The variables
$Y_{I}$ and $Z^I$ carry $c=0$ because of cancellation between the
bosons and fermions.)  Although tree diagrams can be extrapolated to
gauge groups with $c\neq 28$ (much like the Virasoro tree amplitude is
well-defined for $d\neq 26$), one expects problems with the $c\neq 28$
current algebras at the loop level.

Physical open string states are described by the $\GL(1)$-neutral
dimension-one primary fields
\eqn{vertex}
{V_\phi = j^k \phi_k(Z)\,,\qquad 
V_f = Y_I f^I(Z)\,,\qquad
V_g = \p Z^I g_I(Z)\,,}
where $\phi_k(Z)$ has zero $\GL(1)$ charge, $f^I(Z)$ has $+1$ $\GL(1)$
charge and satisfies $\p_I f^I=0$, and $g_I(Z)$ has $-1$ $\GL(1)$
charge and satisfies $Z^I g_I=0$. Through the Penrose transform,
$\phi_k(Z)$ describes $\N=4$ super-Yang-Mills states and $f^I(Z)$ and
$g_I(Z)$ describe $\N=4$ conformal supergravity states whose twistor
description will be explained elsewhere.  Note that the conformal
supergravity states contribute poles to loop amplitudes and multitrace
tree amplitudes, but do not contribute to tree amplitudes involving a
single trace where all external states are super-Yang-Mills
states. This is because all intermediate states in single-trace tree
amplitudes must transform in the adjoint representation of the current
algebra. So for computing tree amplitudes involving a single trace,
the conformal supergravity states can be ignored.

\subsection{Tree amplitudes}

$N$-point tree-level scattering amplitudes are computed in this
string theory by the formula
\eqn{amplitude}
{A = \sum_{d=0}^{N-2} \left\langle c V_1 (z_1) c V_2 (z_2) c V_3(z_3) \int
  \de z_4 V_4(z_4) \cdots \int \de z_N V_N(z_N) \right\rangle_d}
where $\langle ~~\rangle_d$ denotes the correlation function on a disk
with instanton number $d$. Note that just as the Euler number $\int
\de^2 z \sqrt{g} R(z)\,$ is a topological quantity constructed from
the worldsheet metric, the instanton number $\int \de^2 z
\,\e^{\mu\nu} F_{\mu\nu}\,$ is a topological quantity constructed from
the worldsheet gauge field.  Since $\int \de^2 z \sqrt{g} R=4\pi\,$ on
a sphere, a gauge field $A_\mu$ with instanton number $\int \de^2 z
\,\e^{\mu\nu} F_{\mu\nu}=2\pi d$ can be identified with the spin
connection $\Gamma_{\mu\nu}{}^\rho$ in conformal gauge by
\eqn{ident}
{A_z = {d\over 2} \Gamma_{zz}{}^z\,,\qquad 
A_{\bar z} = {d\over 2} ~\Gamma_{\bar z\bar z}{}^{\bar z}\,.}
So for instanton number $d$, the worldsheet action $S=\int \de^2
z\, (Y_{L I} \nabla_{\bar z} Z^I_L + Y_{R I} \nabla_{z} Z^I_R )$
describes worldsheet fields whose conformal weights are twisted by
$-{d\over 2}$ times their $\GL(1)$ charge, i.e.\ $(Y_I,Z^I)$ carries
conformal weight $(1+{d\over 2}, -{d\over 2})$.

On a disk with instanton-number $d$, functional integration over the
zero modes naively gives the measure factor
\eqn{measureone}
{\langle c \p c \p^2 c  v \Phi(Z)\rangle_d =\int 
\de^{8+8d} Z\, \Phi(Z)}
where each $Z^I$ has $d+1$ zero modes. But since the vertex operators
for super-Yang-Mills states are $\GL(1)$-neutral and independent of
the $v$ ghost, this naively gives zero times $ \infty$ where the
$\infty$ comes from writing $Z^I = \hat Z^I r$ and integrating over
the scale factor $r$. One way to regularize~\eqref{measureone} is to
insert the BRST-invariant operator
\eqn{Rdef}
{R(z) = v(z) \d(r(z)-1) + c(z) \p r(z) \d(r(z)-1)}
into the correlation function. Since $\p R= Q (\p r \d(r-1) )$ is
BRST-trivial, it is irrelevant where $R$ is inserted on the worldsheet
when all external states are on-shell. However, since we will later
want to describe a string field theory action involving off-shell
states, it is more convenient to regularize~\eqref{measureone} by
working in a ``small'' Hilbert space where off-shell states are
required to be $\GL(1)$-neutral and to be independent of the $v$ zero
mode. Functional integration in this ``small'' Hilbert space is
defined by
\eqn{measuretwo}
{\langle c \p c \p^2 c  \Phi(Z)\rangle_d =\int Z 
\de^{7+8d} Z\,\Phi(Z)}

{\renewcommand\belowcaptionskip{-2em}
\EPSFIGURE{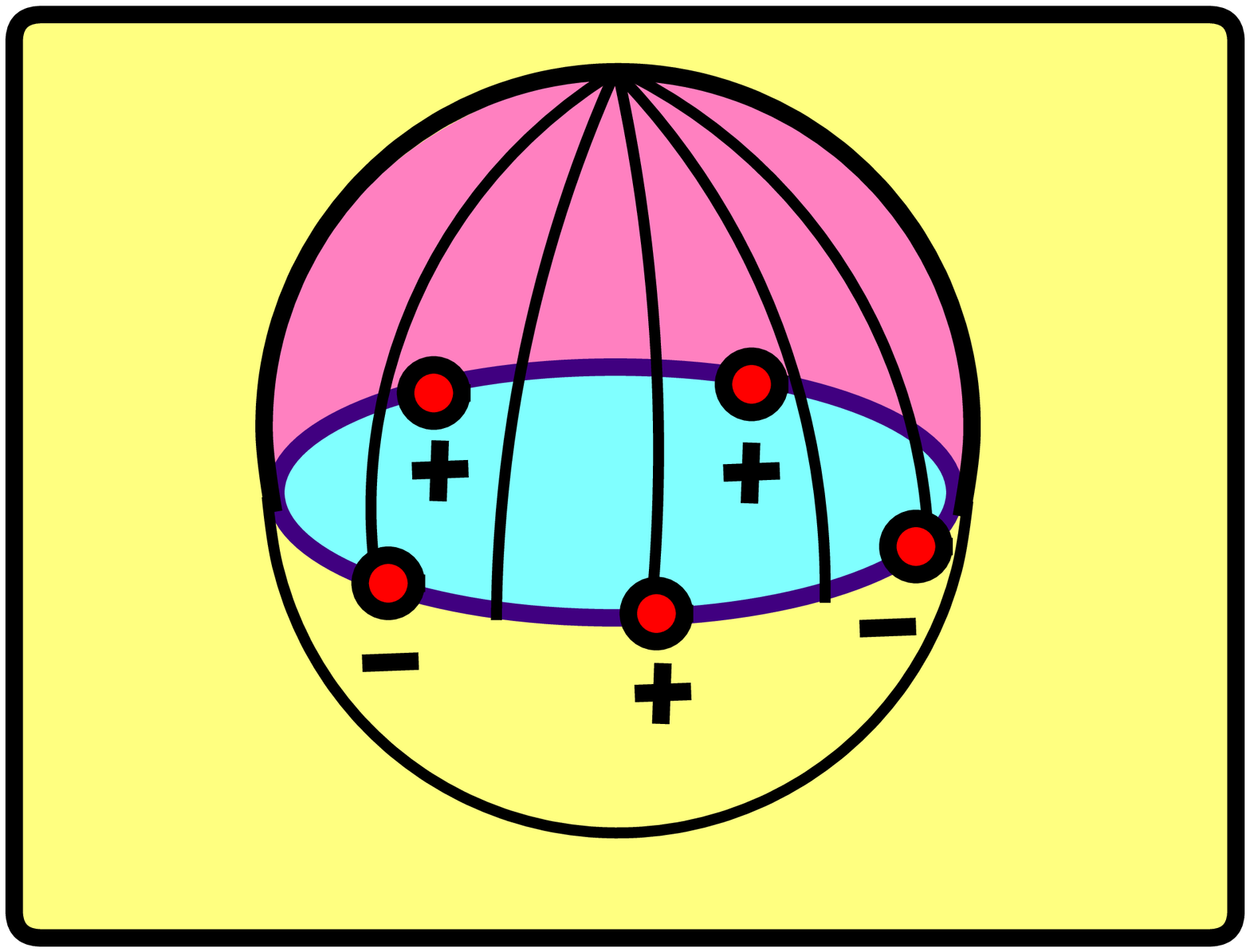,width=.4\textwidth}{The Northern hemisphere
  represents a disk diagram. In this example, five open string vertex
  operators are inserted at its boundary. Unlike the models
  of~\cite{wt} and~\cite{nejvafa}, the twistor variables in this model
  are real.\label{hemipic}}
}

{\sloppy \noindent where $\int Z \de^7 Z$ denotes integration over the
projective space $RP^{3|4}$. For on-shell external states, one can
easily check that this definition is equivalent to inserting the
operator~\eqref{Rdef} into~\eqref{measureone}.

}

As was shown in~\cite{berkovits}, the $N$-point tree amplitudes
computed using the prescription of~\eqref{amplitude} reproduce the
maximal degree D-instanton contribution in the model of~\cite{wt}, and
have been conjectured to reproduce $N$-point super-Yang-Mills tree
amplitudes.  Yang-Mills tree amplitudes are computed as stringy disk
diagrams, such as the one in figure~\ref{hemipic}, using the
Yang-Mills vertex operators
\begin{eqnarray}
V_r(z_r) &=&  j^k(z_r) 
\d \left({{\l^2(z_r)}\over{\l^1(z_r)}}-
{{\pi^2_r}\over{\pi^1_r}}\right)\times
\label{compv}\\&&
\times \exp \biggl( i{{\mu^\ad(z_r)}\over{\l^1(z_r)}}\bar\pi_{r\ad}
\pi_r^1\biggr) \phi_{r k}\biggl({{\psi^A(z_r) \pi_r^1}
  \over{\l^1(z_r)}}\biggr),
\nonumber
\end{eqnarray}
where the external momentum is $p_r^{a\ad} = \pi_r^a \bar\pi_r^\ad$,
$\pi_r^a$ and $\bar\pi_r^\ad$ are independent real quantities in
signature $(2,2)$,
\eqn{phikdef}
{\phi_{ k}\left({{\psi^A \pi^1}\over{\l^1}}\right) = (\pi^1)^{-2}
  \left[ A_{+ k} + \left({{\psi^A \pi^1}\over{\l^1}}\right)^4 A_{- k}
    \right], }
and the Yang-Mills gauge field is 
$$
A_{a\ad k} = \pi_a s_\ad A_{+ k} + \bar s_a \bar\pi_\ad A_{- k}
$$
where $s_\ad$ and $\bar s_a$ are defined such that
$\pi^a \bar s_a = 1$ and $\bar\pi^\ad s_\ad=1$.

For example, the three-point amplitude gets contributions from the
degree zero and degree one correlation functions
\eqn{cubamp}
{A = \langle   c V_1(z_1) c V_2(z_2) c V_3(z_3)\rangle_{d=0}
+ \langle  c V_1(z_1) c V_2(z_2) c V_3(z_3)\rangle_{d=1}\,.}
The $d=0$ piece contributes to the $(++-)$ amplitude while the $d=1$
term contributes to the $(--+)$ amplitude where the signs indicate
helicities.  At degree zero, $Z^I(z)$ has zero conformal weight with
constant zero modes, i.e.
\eqn{zerod}
{\l^a = a^a\,,\qquad
\mu^\ad = b^\ad\,, \qquad 
\psi^A = \g^A\,.}
Using $\GL(1)$ invariance to gauge $a^1=1$, one obtains
\begin{eqnarray}
\left\langle \prod_{r=1}^3 c V_r(z_r)\right\rangle_{d=0} &=& \int \de
a^2 \int \de^2 b^\ad \int \de^4 \g^A 
\d\left( a^2 -{\pi^2_1\over\pi^1_1}\right)
\d\left( a^2 -{\pi^2_2\over\pi^1_2}\right)
\d\left( a^2 -{\pi^2_3\over\pi^1_3}\right)\times
\nonumber\\&&
\times
\exp \left(i b^\ad \sum_{r=1}^3 \bar\pi_{r\ad}\pi_r^1 \right)
f^{k_1 k_2 k_3}\prod_{r=1}^3 \phi_{r k_r} (\g^A \pi_r^1)
\nonumber\\
&=&
\d\left( {\pi^2_1\over \pi_1^1} -{\pi^2_2\over\pi^1_2}\right)
\d\left( {\pi^2_1\over \pi_1^1} -{\pi^2_3\over\pi^1_3}\right)
\d^2\left(\sum_{r=1}^3 \bar\pi_{r\ad}\pi_r^1 \right) \times 
\nonumber\\&& 
\times (\pi_1^1\pi_2^1\pi_3^1)^{-2} \Tr\Bigl( (\pi_3^1)^4 [A_{+ 1},
  A_{+ 2}] A_{- 3} + (\pi_1^1)^4 [A_{+ 2}, A_{+ 3}] A_{- 1} +
\nonumber\\&&
           \hphantom{\times (\pi_1^1\pi_2^1\pi_3^1)^{-2} \Tr\Bigl(}\!
+(\pi_2^1)^4 [A_{+ 3}, A_{+ 1}] A_{- 2} \Bigr)\,.
\label{cubzeroo}
\end{eqnarray}

After scaling $\pi_r^a$ and $\bar\pi_r^\ad$ in opposite directions
until $\pi_1^1=\pi_2^1=\pi_3^1$,~\eqref{cubzeroo} can be written~as
\begin{eqnarray}
\left\langle \prod_{r=1}^3 c V_r(z_r)\right\rangle_{d=0} &=& 
\d^4\left( \sum_{r=1}^3 \bar\pi_{r\ad}\pi_{r a}\right)
(\bar\pi_1^\ad \bar\pi_{2 \ad}) \times 
\nonumber\\&&
\times \Tr ( [A_{+ 1}, A_{+ 2}] A_{- 3} + [A_{+ 2}, A_{+ 3}] A_{- 1} +
       [A_{+ 3}, A_{+ 1}] A_{- 2} )\,.\qquad
\label{cubzer}
\end{eqnarray}
Note that momentum conservation implies that $(\bar\pi_1^\ad
\bar\pi_{2 \ad}) = (\bar\pi_2^\ad \bar\pi_{3 \ad}) = (\bar\pi_3^\ad
\bar\pi_{1 \ad}) $ when $\pi_1^1=\pi_2^1=\pi_3^1$.

At degree one, 
$Z^I$ has $-1/2$ conformal weight so it has the zero modes
\eqn{zeromtwo}
{\l^a(z) = a^a + \tilde a^a z\,,\qquad
\mu^\ad(z) = b^\ad +\tilde b^\ad z\,,\qquad 
\psi^A(z) = \g^A + \tilde\g^A z\,.}
It is convenient to describe these zero modes using the variables
$[u_r, b^{a\ad}, \g^{a A}]$ where
\eqn{newvar}
{u_r = {{a^2 +\tilde a^2 z_r} \over{a^1 + \tilde a^1 z_r}}\,,\qquad  
b^\ad = a_a b^{a\ad}\,,\qquad
\tilde b^\ad =\tilde a_a b^{a\ad},\quad
\g^A = a_a \g^{a A}\,,\qquad
\tilde \g^A = \tilde a_a \g^{a A}\,.}
After gauging $a^1=1$ using $\GL(1)$ invariance, the jacobian from
going to $(a^2, \tilde a^a,$ $ b^\ad,\tilde b^\ad ,$ $ \g^A,\tilde\g^A)$
variables to $(u_r, b^{a\ad},\g^{a A})$ variables is
$(u_1-u_2)(u_2-u_3)(u_3-u_1)$.  So one finds
\begin{eqnarray}
\left\langle \prod_{r=1}^3 c V_r(z_r)\right\rangle_{d=1} &=& \int
\de^3 u_r \int \de^4 b^{a\ad} \int \de^8 \g^{a A} (u_1 - u_2)^{-1}
(u_2-u_3)^{-1} (u_3-u_1)^{-1}\times
\nonumber\\&&
\times\d\left( u_1 -{\pi^2_1\over\pi^1_1}\right) \d\left( u_2
-{\pi^2_2\over\pi^1_2}\right) \d\left( u_3
-{\pi^2_3\over\pi^1_3}\right) \times
\nonumber\\&&
\times\exp \left(i b^{a\ad} \sum_{r=1}^3
\bar\pi_{r\ad}\pi_{r a} \right) f^{k_1 k_2 k_3}\prod_{r=1}^3 \phi_{r
  k_r} (\pi_{r a}\g^{a A})
\nonumber\\
&=& \d^4\left( \sum_{r=1}^3 \bar\pi_{r\ad}\pi_{r a}\right) (\pi_1^a
\pi_{2 a})^{-1} (\pi_2^b \pi_{3 b})^{-1}(\pi_3^c\pi_{1 c})^{-1} \times
\nonumber\\&&
\times \Tr\Bigl( (\pi_1^d\pi_{2 d})^4 [A_{- 1}, A_{- 2}] A_{+ 3} +
(\pi_2^d\pi_{3 d})^4 [A_{- 2}, A_{- 3}] A_{+ 1} +
\nonumber\\&&
            \hphantom{\times \Tr\Bigl(}\!
+(\pi_3^d\pi_{1 d})^4 [A_{- 3}, A_{- 1}] A_{+ 2} \Bigr)\,.
\label{cubone}
\end{eqnarray}
If one scales $\pi_r^a$ and $\bar\pi_r^\ad$ in opposite directions
until $\bar\pi_1^{\dot 1}=\bar\pi_2^{\dot 1}=\bar\pi_3^{\dot 1}$,
conservation of momentum implies that $(\pi_1^a \pi_{2 a}) = (\pi_2^\a
\pi_{3 a}) = (\pi_3^\a\pi_{1 a}) $ and~\eqref{cubone} reduces to the
parity conjugate of~\eqref{cubzer}.  One can easily check that the sum
of the degree zero and degree one contributions in~\eqref{cubzer}
and~\eqref{cubone} correctly reproduce the three-point Yang-Mills
couplings.\footnote{Since
the trace over group theory factors in this string theory comes from
current algebra OPE's and not from Chan-Paton factors, changing the
order of the vertex operators on the boundary does not alter the order in
the trace. This implies that the degree zero
correlation function $\langle cV_1(z_1) cV_2(z_2) cV_3(z_3) \rangle_{d=0}$
contributes with opposite sign from the correlation function
$\langle cV_2(z_1) cV_1(z_2) cV_3(z_3) \rangle_{d=0}$.
So these different cyclic orderings naively cancel each
other, which would imply that the $d=0$ term does not contribute to
the on-shell three-point amplitude~\cite{wittensign}.
To get a non-vanishing $d=0$ contribution, one should define an
analytic continuation so that switching the order of the vertex
operators does not switch the sign. This can be accomplished by
multiplying the $d=0$ correlation function by the factor
$\mathrm{sign}(\bar\pi_1^{\dot a}\bar\pi_{2 \dot a})$. Note that under
parity symmetry, this means the $d=1$ correlation function should be
multiplied by $\mathrm{sign}(\pi_1^{a}\pi_{2 a})$, which is necessary
for converting the standard Jacobian $|(u_1-u_2)(u_2-u_3)(u_3-u_1)|$
to the holomorphic Jacobian $(u_1-u_2)(u_2-u_3)(u_3-u_1)$ which was
used in the degree one computation.}

For $N$-point tree amplitudes, the formula is
\begin{eqnarray}
A(\lambda_i,\bar \lambda_i,\psi_i)&=&
\frac{\int\de^{2d+2} a \,\de^{2d+2} b \, \de^{4d+4}\gamma
\int \de z_1 \cdots\int \de z_N}{\mathrm{Vol(GL(2))}}\times
\nonumber\\&&
\times\prod_{r=1}^{N}
\frac{1}{(z_r -z_{r+1\mathrm{\,\,mod\,\,}N})} 
\prod_{r=1}^N \d \left({{\l^2(z_r)}\over 
{\l^1(z_r)}} -
{\pi_r^2\over\pi_r^1}  \right)
\exp \left( i
{{\mu^\ad(z_r)}\over{\l^1(z_r)}} 
\bar\pi_{r\ad}\pi^1_r\right)\times\qquad
\nonumber\\&&
\times \Tr
\left[\phi_1\left({{\psi^A(z_1)\pi_1^1}\over{\l^1(z_1)}}\right)
  \phi_2\left({{\psi^A(z_2)\pi_2^1}\over{\l^1(z_2)}}\right) \cdots
  \phi_N\left({{\psi^A(z_N)\pi_N^1} \over{\l^1(z_N)}}\right)\right]
\end{eqnarray}
where
$$
\l^a(z) = \sum_{k=0}^d a_k^a z^k\,,\qquad
\mu^\ad(z) = \sum_{k=0}^d b_k^\ad z^k\,,\qquad
\psi^A(z) = \sum_{k=0}^d \gamma_k^A z^k\,,
$$
$(a_k^a, b_k^\ad,\gamma_k^A)$ are the zero modes of $Z^I$ on a disk,
and the $\SL(2)$ part of $\GL(2)$ can be used to fix three of the
$z_r$ integrals and reproduce the $(b,c)$ correlation function.  When
$d=1$ and $d=N-3$, this formula has been verified to give the correct
super-Yang-Mills MHV and googly tree amplitudes. Since the above
formula will be obtained from the string field theory action of this
paper (which is expected to have unitary factorization properties), we
consider this strong evidence that the formula gives the correct
super-Yang-Mills tree amplitudes for arbitrary helicity-violation.

\section{Cubic string field theory action}\label{csft}

\subsection{Kinetic term}

The first step in constructing a field theory action is to construct a
kinetic term whose equation of motion and gauge invariance describe
the physical spectrum. The off-shell string field will be described by
the wave functional
\eqn{Phidef}
{|\Phi\rangle =  \Phi[Y,Z,j,b,c,u,v] |0\rangle}
where $|0\rangle$ is a ground state satisfying
\eqn{ground}
{Z_n^I|0\rangle = 
Y_{(n-1) I}|0\rangle =  j^k_{n-1}|0\rangle = 
b_{n-2}|0\rangle =  c_{n+1}|0\rangle = 
u_{n-1}|0\rangle =  v_{n}|0\rangle =  0 \qquad\hbox{for~}  
n>0\,.}
We are using the usual mode expansion
for open string worldsheet variables on a strip with 
$0\leq \sigma\leq \pi$, e.g.
\begin{equation}
\begin{array}[b]{rclrcl}
Z_{L}^I(\tau,\sigma) &=&\displaystyle \sum_{n=-\infty}^\infty Z_n^I
  e^{i n(\tau -\sigma)}\,,\qquad&
Z_{R}^I (\tau,\sigma) &=&\displaystyle \sum_{n=-\infty}^\infty Z_n^I
e^{i n(\tau +\sigma)}\,,
\\[10pt]
Y_{L I} (\tau,\sigma) &=&\displaystyle \sum_{n=-\infty}^\infty Y_{nI}
e^{i n(\tau -\sigma)}\,,\qquad&
Y_{R I}(\tau,\sigma) &=&\displaystyle \sum_{n=-\infty}^\infty Y_{nI}
e^{i n(\tau +\sigma)}\,,
\end{array}
\end{equation}
where $(Z_n^I)^\dagger = Z_{-n}^I$, $(Y_{n I})^\dagger = Y_{-n I}$,
and $Y_{n I} Z^J_m - (-1)^{\mathrm{sign}(I)} Z^J_m Y_{n I}= \d_{m+n}
\d_I^J$.

In addition to the usual requirement that the string field
$|\Phi\rangle$ carries $+1$ ghost number, it will also be required
that $|\Phi\rangle$ is in the ``small'' Hilbert space defined in
section~\ref{reviewed}. In other words, $|\Phi\rangle$ must be
$\GL(1)$-neutral and independent of the $v$ ghost zero mode, i.e.\
\eqn{newcond}
{J_0 |\Phi\rangle =0
\qquad\hbox{and} \qquad
u_0 |\Phi\rangle =0\,.}
However, note that $\Phi$ is allowed to depend on inverse powers of
$Z_0^I$ since this dependence is necessary for describing the on-shell
twistor wavefunctions for super-Yang-Mills states.  So the generic
off-shell string field is
\eqn{generic}
{|\Phi\rangle=\sum_s\phi_s(Z_0) f_s(Z_{-n}^I, Y_{-n I}, c_{2-n},
  b_{-1-n}, j^k_{-n}, v_{-n}, u_{-n})|0\rangle}
where $f_s$ is an arbitrary polynomial in oscillators $(Z_{-n}^I,
Y_{-n I}, c_{2-n}, b_{-1-n}, j^k_{-n}, v_{-n}, u_{-n})$ for $n>0$ such
that its $\GL(1)$ charge cancels the $\GL(1)$ charge of $\phi_s(Z_0)$.

Under the above conditions, one can define a kinetic term as
\eqn{kinetic}
{S_{\mathrm {kin}} = \langle \Phi | Q |\Phi\rangle}
where $Q$ is the BRST operator defined in~\eqref{brst} and $\langle
\Phi |=\langle 0| \Phi^\dagger$ is obtained from $|\Phi\rangle$ by
hermitean conjugation which switches the signs of all mode indices.
Note that the BPZ conjugate bra-vacuum $\langle 0|$ is defined to
satisfy
\eqn{groundl}
{\langle 0|Z_n^I = 
\langle 0|Y_{(n+1) I} =  \langle 0|j^k_{n+1} = 
\langle 0|b_{n+2} = \langle 0| c_{n-1} = 
\langle 0|u_{n+1} = \langle 0| v_{n} =  0\qquad \hbox{for~}  n<0\,.}
As usual, one can also write the kinetic term in~\eqref{kinetic} as
$\int \Phi * Q\Phi$ where the functional integral is over all
configurations of the first half of the string that is identified with
the second half of the string.

The zero mode normalization of the action will be defined
in the ``small'' Hilbert space~by 
\eqn{zeromode}
{\langle 0| c_{-1} c_0 c_1\,\phi(Z_0) |0\rangle = 
\int Z_0 \de^7 Z_0\, \phi(Z_0)}
where $\int Z_0 \de^7 Z_0$ is an integral over $RP^{3|4}$ and $\langle
0|$ is the BPZ conjugate of $\ket 0$.  Note that if one tried to
define the zero mode normalization of the action in the ``large''
Hilbert space involving the $v$ zero mode and the $\GL(1)$ scale
factor, one would run into the problem that the kinetic term
of~\eqref{kinetic} should have ghost-number four. Since $Q$ has ghost
number one, this would mean that $|\Phi\rangle$ must carry
half-integer ghost number.

This situation has an analog in bosonic 
closed string field theory. Naively, the zero mode normalization
in bosonic closed string field theory should be 
\eqn{bsft}
{\langle 0| c_{-1} c_0 c_1 \bar c_{-1} \bar c_0 \bar c_1 |0\rangle =1}
where $c$ and $\bar c$ are the left and right-moving Virasoro ghosts.
However, this would mean that the closed string field $|\Phi\rangle$
in the action $\langle\Phi |Q|\Phi\rangle$ should carry half-integer
ghost number.  The solution is to work in a ``small'' Hilbert space
where $|\Phi\rangle$ is restricted to be independent of the $(c-\bar
c)$ zero mode and to be neutral under $(L_0 - \bar L_0)$ rotations,
i.e.\ $(b_0 -\bar b_0) |\Phi\rangle = (L_0-\bar L_0)|\Phi\rangle=0$.
Note that these restrictions on the bosonic closed string field are
analogous to the restrictions of~\eqref{newcond} for the twistorial
open string field.  In this ``small'' Hilbert space, one can define
the zero mode normalization as
\eqn{bsfttwo}
{\langle 0| c_{-1} c_1 (c_0 +\bar c_0) \bar c_{-1} \bar c_1 |0\rangle
  =1}
so that $\langle\Phi |Q|\Phi\rangle$ is nonvanishing when the closed
string field $|\Phi\rangle$ carries $+2$ ghost number.

The kinetic term of~\eqref{kinetic} implies the equations of motion
$Q|\Phi\rangle=0$ and the gauge invariance $\d|\Phi\rangle =
Q|\Omega\rangle$ where $J_0|\Omega\rangle =
u_0|\Omega\rangle=0$. These equations are consistent since
$[Q,u_0]=J_0$ and $[Q,J_0]=0$.  One can check that the only states in
the cohomology of $Q$ at ghost-number one which satisfy
$u_0|\Phi\rangle = J_0 |\Phi\rangle = 0$ are
\eqn{pvertex}
{c_1 j^k_{-1} \phi_k(Z_0) |0\rangle\,,\qquad 
c_1 Y_{-1 I} f^I(Z_0) |0\rangle\,,\qquad
c_1 Z^I_{-1} g_I(Z_0) |0\rangle\,,}
which correspond to the super-Yang-Mills and conformal supergravity
vertex operators of~\eqref{vertex}.  Although the kinetic term
$\langle \Phi|Q|\Phi\rangle$ reproduces the super-Yang-Mills and
conformal supergravity spectrum, the equations of motion coming from
$Q|\Phi\rangle =0$ are completely different from the standard
super-Yang-Mills and conformal supergravity equations of motion. For
example, since the only constraint on the super-Yang-Mills twistor
field $\phi_k(Z_0)$ is $\GL(1)$-invariance, it does not even appear in
the $\langle \Phi|Q|\Phi\rangle$ kinetic term. However, $\phi_k(Z_0)$
will appear in the cubic interaction term.

\subsection{Cubic term: the $d=0$ part}

The cubic term in the string field theory action can be determined by
the requirements that it preserves a nonlinear version of the gauge
invariance $\d |\Phi\rangle = Q|\Omega\rangle$ and that it reproduces
the desired on-shell three-point amplitudes. Since the three-point
amplitude involves correlation functions of degree zero and degree
one, we will need two cubic terms in the field theory action.

The cubic term of degree zero is easily obtained by using Witten's
star product~\cite{wittensft} to glue the left and right halves of two
string fields to give a third string field $|\Phi\rangle *
|\Phi\rangle$.  The cubic term of degree zero is
\eqn{cubiczero}
{S_{d=0} = g \langle \Phi| (|\Phi\rangle * |\Phi\rangle)
= g \int \Phi * \Phi * \Phi \,.}
If one does not include the cubic term of degree one, one would have
the action
\eqn{cubiczerot}
{S = \langle \Phi| Q|\Phi\rangle + {2\over 3} g \langle \Phi|
  (|\Phi\rangle * |\Phi\rangle)\,,}
which has the nonlinear gauge invariance
\eqn{nonlint}
{\d|\Phi\rangle = Q|\Lambda \rangle+ g(|\Phi \rangle *|\Lambda
  \rangle- |\Lambda \rangle* |\Phi\rangle)}
and describes the version of $\N=4$ super-Yang-Mills with only
self-dual interactions~\cite{siegelym}.

Note that~\eqref{cubiczerot} is gauge invariant using the usual
axioms of open string field theory. Namely, $Q$ is a derivation
with respect to the star product, i.e.
\eqn{leibniz}
{Q(\ket{\Phi_1} * \ket{\Phi_2}) = (Q\ket{\Phi_1})* \ket{\Phi_2} -
  \ket{\Phi_1} * (Q\ket{\Phi_2})}
where the minus sign is because $|\Phi\rangle$ is fermionic,
$\int Q\Phi =0$, $Q$ is nilpotent, and the star product is associative,
i.e.
\eqn{associ}
{(\Phi_1 * \Phi_2)*\Phi_3 = \Phi_1* (\Phi_2 * \Phi_3)\,.}
Since $u_0$ and $J_0$ are integrals of dimension-one currents, they
also act as derivatives with respect to the star product. So the
constraints $u_0|\Phi\rangle = J_0|\Phi\rangle=0$ are preserved by the
gauge transformation of~\eqref{nonlint} if $u_0|\Lambda\rangle =
J_0|\Lambda\rangle=0$.

\EPSFIGURE[t]{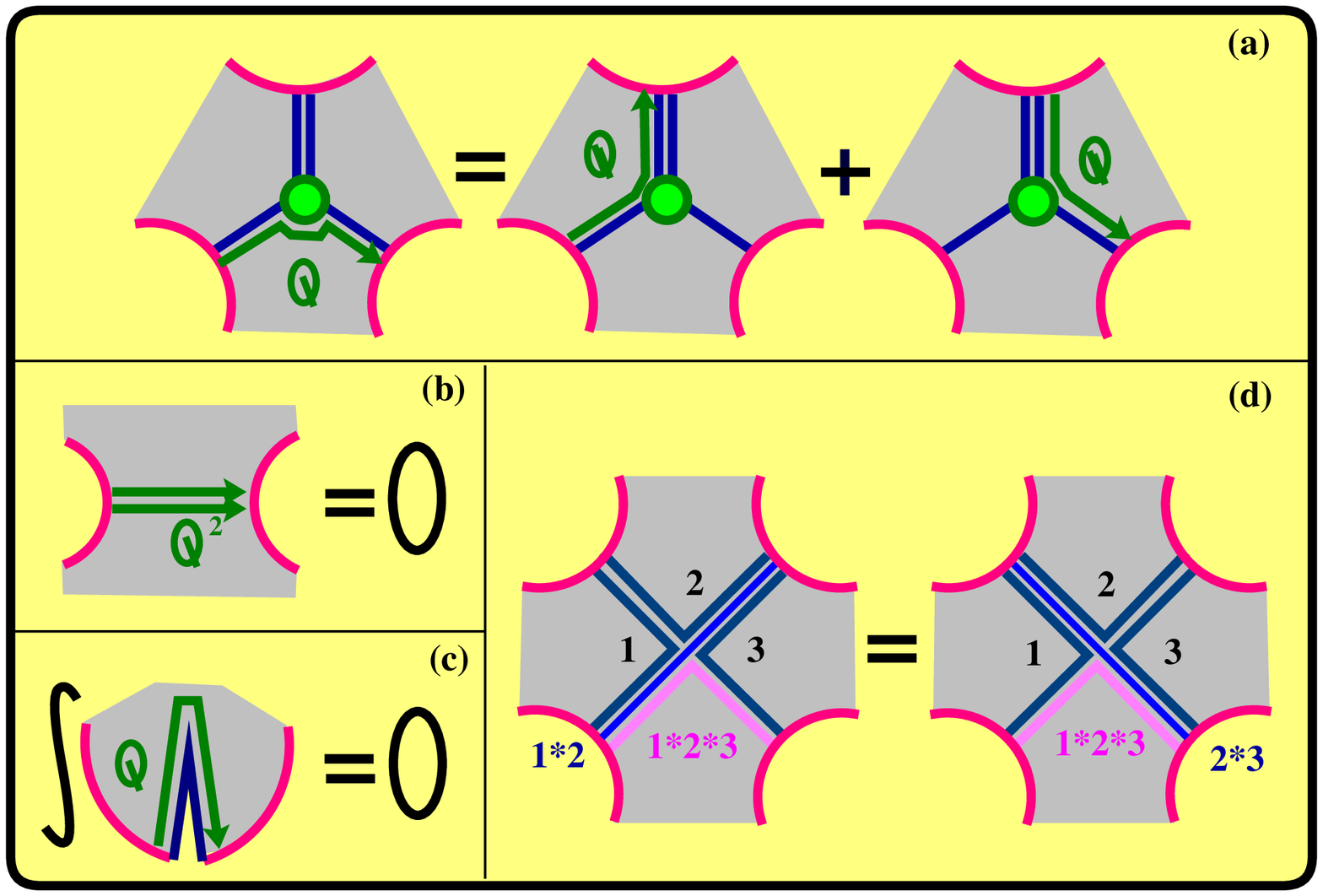,width=.8\textwidth}{Diagrams representing the axioms
  of string field theory: ($a$) $Q$ is a derivation of the
  star-product. Note that $u_0$ and $J_0$ are also derivations of the
  star-product and that a BRST-invariant insertion $F(\pi/2)$ at the
  midpoint does not spoil the identity.  ($b$) Nilpotency of $Q$.
  ($c$) $\int Q\Phi =0$.  ($d$) Associativity of the
  star-product.\label{axiomspic}}

\subsection{Cubic term: the $d=1$ part}

To construct the cubic term of degree one, first note that one can
define a BRST-invariant spectral flow operator~\cite{NB, Lech,
  Lechsiegel}
\eqn{spectral}
{F(z) = e^{i\s(z)} = \exp\left(i\int^z \de y\, Y_I(y) Z^I(y)\right) (1-
  i c(z) u(z))}
where $\p\s = \{Q, u\} = Y_I Z^I - \p(cu)$ is the total $\GL(1)$
current.  Note that $\s(y)\s(z)$ has no singularity and that $F(z)$
can be expressed in operator language as
\eqn{spectraly}
{F(z)= \d^8 (Y(z)) (1 - i c(z) u(z))\,.}

Since the $\GL(1)$ gauge field $A_{\bar z}$ couples to $\p_z\s$ in the
worldsheet action, the correlation function on a disk of instanton
number $d$ is equivalent to the correlation function on a disk of
instanton number zero with $d$ spectral flow operator insertions.  To
see this, suppose that $\p_z A_{\bar z}= \sum_{r=1}^d \d^2 (z-z_r)$ so
that the worldsheet field strength is concentrated at points on the
worldsheet.  Exponentiating the term $-i\int \de^2 z (A_{\bar z} J_z)$
in $i S_{\mathrm{worldsheet}}$ therefore gives the contribution
$\prod_{r=1}^d e^{i\s(z_r)}$.

These spectral flow insertions give a background charge to the
worldsheet variables which can be mimicked by twisting their conformal
weight according to their $\GL(1)$ charge.  One can think of $\d^8
(Y(z))$ as forcing $Y(z)=0$ at the insertion point $z$; $Y(y)$ is then
proportional to $(y-z)$ near this point, and therefore the dual
variable $Z(y)$ is allowed to blow up as $(y-z)^{-1}$ near $z$. The
field $Z$ therefore has one new zero mode on the disk, and it can
describe a curve of a higher degree.

When all external states are on-shell, the locations of these
insertions are irrelevant since $\p F = Q(i u e^{i\s})$ is BRST-exact.
But in open string field theory, the external states are off-shell, so
the locations of these spectral flow insertions are relevant. However,
the unique location for these insertions which preserves gauge
invariance is the midpoint of the string.  So the cubic term of degree
one in the action will be defined as
\eqn{cubicone}
{S_{d=1} = g' \langle \Phi| F\left(\frac{\pi}{2}\right) (|\Phi\rangle
  * |\Phi\rangle) = g'\int F\left(\frac{\pi}{2}\right)\Phi * \Phi *
  \Phi\,.}
The extra insertion of $F(\pi/2)$ does not spoil BRST invariance since
$[Q, F(\pi/2)]=0$.

Since midpoint insertions cause problems~\cite{wendt,aref} in cubic
open superstring field theory~\cite{wittensuper}, one might be worried
that similar problems could arise here. Fortunately, this does not
occur.  Unlike the picture-raising operators in the RNS formalism,
$F(y)$ has no singularities with $F(z)$ so these insertions do not
cause contact term divergences when the midpoints
collide~\cite{wendt}.  Also, since the kinetic term in the action does
not require midpoint insertions, there is no need to truncate out
states which are annihilated by the spectral flow
operator~\cite{aref}.

\EPSFIGURE[t]{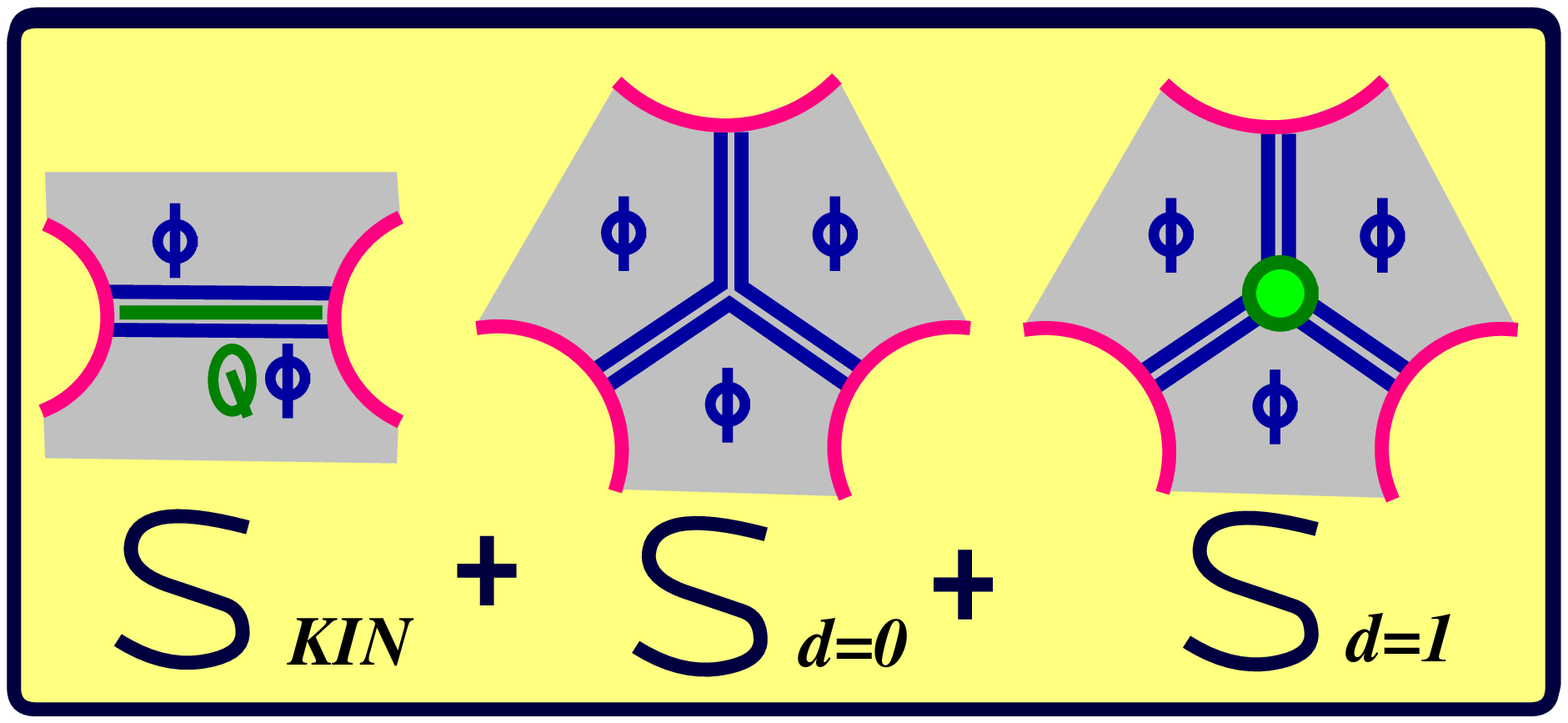,width=90mm}{A visual representation of the total
  twistorial string field theory action.\label{actionpic}}

So the complete open string field theory action is (see also
figure~\ref{actionpic})
\eqn{complete}
{S=\langle \Phi| Q|\Phi\rangle +{2\over 3}
 g \langle \Phi| (|\Phi\rangle * |\Phi\rangle)
+ {2\over 3} g' \langle \Phi| F\left(\frac{\pi}{2}\right)
(|\Phi\rangle * |\Phi\rangle)\,,}
which has the nonlinear gauge invariance
\eqn{nonlinc}
{\d|\Phi\rangle = Q|\Lambda \rangle+ g(|\Phi \rangle *|\Lambda
  \rangle- |\Lambda \rangle* |\Phi\rangle) +
  g'F\left(\frac{\pi}{2}\right) (|\Phi \rangle *|\Lambda \rangle-
  |\Lambda \rangle* |\Phi\rangle)\,.}

This action can be put in a more conventional form by first rescaling
the fermionic components of $Z^I$ and $Y_I$ as $\psi^A \to ({g' /
  g})^{1\over 4} \psi^A$ and $\bar\psi_A \to ({g / g'})^{1\over 4}
\bar\psi_A$.  Since the norm and $F(\pi/2)$ rescale by a factor of
$(g/g')$, the action becomes
\eqn{actthree}
{S={g\over { g'}}\langle \Phi| Q|\Phi\rangle + {{2 g^2}\over {3g'}}
  \langle \Phi| (|\Phi\rangle * |\Phi\rangle) + {{2g^2}\over {3g'}}
  \langle \Phi| F\left(\frac{\pi}{2}\right) (|\Phi\rangle *
  |\Phi\rangle)\,.}
Now by rescaling $|\Phi\rangle \to  g^{-1} |\Phi\rangle$, one
obtains the action 
\eqn{actfour}
{S={1\over {g g'}}\left[\langle \Phi| Q|\Phi\rangle + {2\over
      3}\langle \Phi| (|\Phi\rangle * |\Phi\rangle) + {2\over
      3}\langle \Phi| F\left(\frac{\pi}{2}\right) (|\Phi\rangle *
    |\Phi\rangle)\right].}

Note that it is only the product $gg'$ of the two coupling constants
that has an invariant meaning. Although the $d=1$ cubic term looks
more unnatural than the $d=0$ cubic term, we will show in
section~\ref{paritysection} that the operation of ``parity'' exchanges
the two cubic terms.  But let us first demonstrate the equivalence of
our string field theory prescription and the first-quantized procedure
described in~\cite{berkovits}.

\subsection{Equivalence with first-quantized prescription}

To define string Feynman diagrams using the action of~\eqref{actfour},
one needs to gauge fix the string field. Since $\{Q,b_0\}= L_0$, this
can be done using the standard Siegel gauge-fixing condition that
$b_0|\Phi\rangle =0$.  In this gauge, one can use standard open string
field theory methods~\cite{gm,gmw,zwiebach} to show that the string
Feynman diagrams cover the moduli space of open string tree
amplitudes.

\EPSFIGURE[t]{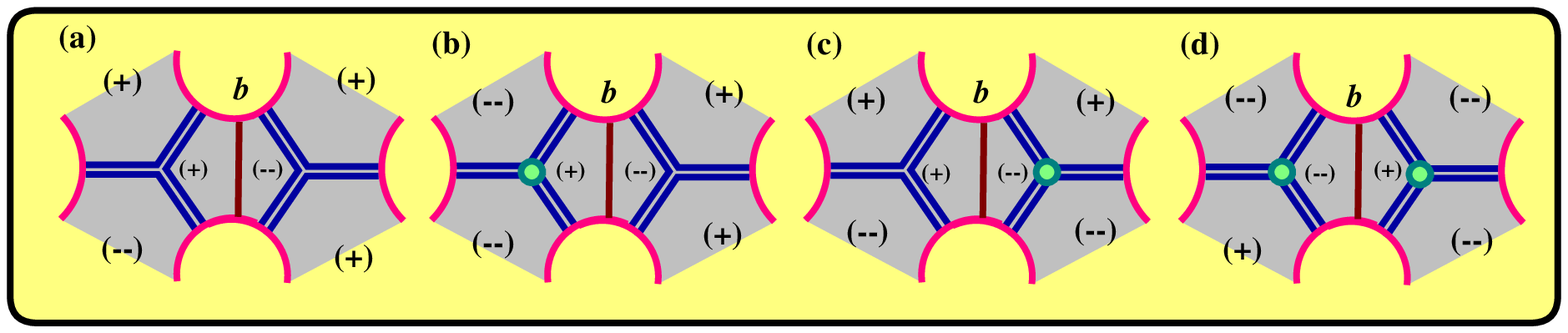,width=\textwidth}{A scattering amplitude of four
  gluons requires two cubic vertices. Each of them is either the $d=0$
  vertex, coupling the $(++-)$ helicities, or the $d=1$ vertex,
  coupling the $(--+)$ helicities.  ($a$) Two $d=0$ vertices,
  contributing to $(+++-)$. ($b$) One $d=0$ and one $d=1$ vertex,
  contributing to $(++--)$. ($c$) One $d=1$ and one $d=0$ vertex,
  contributing to $(--++)$.  ($d$) Two $d=1$ vertices, contributing to
  $(---+)$.  The exact ordering of the external helicities may
  differ. The brown line crossing the intermediate string represents
  an integral of the Virasoro antighost
  $b(\sigma)$.\label{quarticpic}}

A new feature here as compared with bosonic open string field theory
is that there are two types of cubic vertices, one which carries
instanton number zero and the other which carries instanton number
one.  For computing $N$-point tree amplitudes of instanton number $d$,
one gets contributions when $(N\!-\!2\!-\!d)$ cubic vertices carry
instanton number zero and $d$ cubic vertices carry instanton number
one.  When all external states are on-shell, the choice of which cubic
vertices are instanton number zero and which are instanton number one
is irrelevant.  Naively, this would imply that the on-shell $N$-point
tree amplitudes of degree $d$ in equation~\eqref{amplitude} are
multiplied by the combinatoric factor ${{(N-2)!}\over{(N-2-d)!~ d!}}$.
However, as will now be explained, this combinatoric factor is
cancelled when one identifies states which are related by the spectral
flow operator.

In our string field theory action, the off-shell string field is
$\Phi|0\rangle$ where $|0\rangle$ is the ground state defined
in~\eqref{ground}. This ground state is related by the spectral flow
operator $F^p$ to other ground states $|p\rangle= F^p |0\rangle$ where
$F^p = e^{ip\sigma}$, $\p\sigma = Y_I Z^I -\p(cu)$, and $|p\rangle$
satisfies the conditions
\begin{eqnarray}
Z_{n+p}^I|p\rangle &=& 
Y_{(n-1-p) I}|p\rangle = 0
\qquad \hbox{for~}  n>0\,,
\nonumber\\
j^k_{n-1}|p\rangle &=& 
b_{n-2}|p\rangle =  c_{n+1}|p\rangle = 
u_{n-1}|p\rangle =  v_{n}|p\rangle =  0\qquad \hbox{for~}  n>0\,.
\label{groundp}
\end{eqnarray}
Note that $Q \Phi|0\rangle=0$ implies that $Q \Phi|p\rangle= Q \Phi
F^p |0\rangle=0$, so the BRST cohomology constructed from the ground
state $|p\rangle$ is isomorphic to the BRST cohomology constructed
from the ground state $|0\rangle$.  In analogy with the RNS
superstring, we will denote states constructed from the ground state
$|p\rangle$ as states with ``picture'' $p$.

When computing tree amplitudes using the string field theory action
of~\eqref{actfour}, one only includes intermediate states in the zero
picture. This is necessary for unitarity since each physical state
should be represented by a unique string field in the BRST cohomology.
But when computing tree amplitudes using the first-quantized
prescription, functional integration over the worldsheet variables
allows all possible intermediate states in all possible pictures. For
tree amplitudes in the RNS superstring, this difference between string
field theory and first-quantized computations has no effect on
scattering amplitudes since picture in the RNS formalism is a
conserved quantity. So in RNS tree amplitudes, the picture of the
intermediate states is completely determined by the picture of the
external states.

But in this twistorial string field theory, picture is not conserved.
Since the cubic vertex of degree one involves an explicit $F$
insertion, cubic interactions can violate picture by either $one$ or
zero.  This causes a difference between string field theory and
first-quantized computations which cancels the combinatoric factor
${{(N-2)!}\over{(N-2-d)!~d!}}$.

For example, consider the four-point amplitude described by figure 4,
and put all external states in the zero picture. In diagram ($a$), the
intermediate state must be constructed from $|0\rangle \langle 0|$
since otherwise one of the two cubic vertices would have
picture-violation different from zero or one. But in diagram ($b$),
the intermediate state could be constructed either from $|0\rangle
\langle 0|$ or from $|\!-\!1\rangle \langle +1|$. In the first case,
the cubic vertex on the left has picture-violation one and the vertex
on the right has picture-violation zero. And in the second case, the
vertex on the left has picture-violation zero and the vertex on the
right has picture-violation one. Similarly, in diagram ($c$), the
intermediate state could be constructed either from $|0\rangle \langle
0|$ or from $|+1\rangle \langle -1|$. And in diagram ($d$), the
intermediate state must be constructed from $|0\rangle \langle 0|$.

For on-shell external states, the two types of intermediate states in
diagrams ($b$) and ($c$) contribute equally. So the string field
theory computation (which only includes the $|0\rangle\langle 0|$
contribution) is half of the first-quantized computation.  As desired,
this factor of half cancels the factor of two coming from the
combinatoric factor ${{(N-2)!}\over{(N-2-d)!~ d!}}$.  One can easily
check that a similar cancellation occurs for all higher-point tree
amplitudes. This is because the number of choices\pagebreak[3] for intermediate
states is always equal to the number of ways that the vertices can
violate picture, which is equal to the number of ways that the $d=0$
and $d=1$ vertices can be distributed. So the combinatoric factors
cancel for any number of external states.

So using the string Feynman diagrams in Siegel gauge, one reproduces
the first-quantized prescription of~\eqref{amplitude} for tree
amplitudes. Furthermore, it should be possible to show that the poles
in the twistorial string Feynman diagrams correspond to physical
states and are consistent with unitarity. Since the string three-point
amplitudes reproduce the correct cubic interactions, this is strong
evidence that the string theory correctly computes the $N$-point
super-Yang-Mills tree amplitudes.

\section{Parity symmetry}\label{paritysection}

One of the characteristic properties of the twistor formalism is that
the (left-right) parity symmetry is not manifest. Under this symmetry,
the positive and negative helicities are interchanged.  While the
amplitudes with mostly $(+)$ helicities are described by curves of a
small degree, the ``googly'' amplitudes with mostly $(-)$ helicities
require us to consider curves of a large degree which are much more
difficult to deal with. Nevertheless, parity is an exact symmetry of
the super-Yang-Mills theory S-matrix, and it should be possible to
prove this symmetry explicitly.

When the amplitudes are converted to the twistor space, one of the
spinors $\lambda^a$ and $\tilde\lambda^{\dot a}$ (usually
$\tilde\lambda^{\dot a}$) must be Fourier-transformed. Had we
transformed the other spinor $\lambda^a$, we would have obtained the
googly description in terms of the dual twistor space.  We can check
that the external wavefunctions in these two dual pictures are
represented by the Fourier transform over all bosonic as well as
fermionic twistor coordinates.  For example, for the Yang-Mills vertex
operator of~\eqref{compv}, the Fourier transform is
\begin{eqnarray}
\tilde V(Y) &=&\int \de^8 Z\, e^{i Y_I Z^I} V(Z) = \int \de^2\l\,
\de^2\mu\, \de^4\psi \, e^{i(\bar\mu_a\l^a + \bar\l_\ad \mu^\ad
  +\bar\psi_A \psi^A)} \times
\nonumber\\[-1pt minus 2pt]&&
\times \, j^k \d \left({{\l^2}\over{\l^1}}-
       {{\pi^2}\over{\pi^1}}\right) \exp
       \left(i{{\mu^\ad}\over{\l^1}}\bar\pi_{\ad} \pi^1\right)
       (\pi^1)^{-2} \left[A_{+k} + \left({{\psi^A\pi^{1}}\over \l^{
	     1}}\right)^4 A_{-k}\right]\qquad
\nonumber\\[-1pt minus 2pt]
&=& j^k \d \left({{\bar\l_{\dot 2}}\over{\bar\l_{\dot 1}}}-
{{\bar\pi_{\dot 2}}\over{\bar\pi_{\dot 1}}}\right) \exp
\left(i{{\bar\mu_a}\over{\bar\l_{\dot 1}}}\pi^{a}
\bar\pi_{\dot 1}\right) ~(\bar\pi_{\dot 1})^{-2}
\left[A_{-k} + \left({{\bar\psi_A\bar\pi_{\dot 1}}\over 
\bar\l_{\dot 1}}\right)^4 
A_{+k}\right],
\label{fourv}
\end{eqnarray}
where $Y_I=(\bar\mu_a,\bar\l_\ad,\bar\psi_A)$.

Comparing $V(Z)$ of~\eqref{compv} with $\tilde V(Y)$ of~\eqref{fourv},
one sees that performing a parity transformation on the states is
equivalent to performing a Fourier transform of the vertex operator
which switches $Z^I$ with $Y_I$.  This is consistent with
superconformal transformations since the parity operation exchanges
fundamental and antifundamental representations of $\PSU(2,2|4)$.
Although the Fourier transformation acts on the function of the zero
modes of $Z^I$, the stringy completion of this operation will involve
the complete interchange of $Z^I$ and its canonical momentum $Y_I$
\eqn{zyflip}
{Z^I(\sigma) \ \leftrightarrow \ Y_I(\sigma)\,,
} 
including the oscillators.  In subsection~\ref{section4.2}, we will be more
precise how the $Z^I$ and $Y_I$ variables are interchanged.

\pagebreak[3]

\subsection{Parity symmetry of on-shell amplitudes}

Before discussing the parity symmetry of the string field theory
action, it will be useful to demonstrate that the on-shell amplitude
prescription of~\eqref{amplitude} is invariant under parity
transformations\footnote {This demonstration was inspired by comments
  of Edward Witten and Warren Siegel on parity symmetry in twistor
  calculations.}.  Suppose one has an $N$-point amplitude involving
$d+1$ negative-helicity gluons and $(N-d-1)$ positive-helicity
gluons. Using the prescription of~\eqref{amplitude}, this is computed
by the correlation function of $N$ vertex operators $V(Z(z_r))$
of~\eqref{compv} on a disk of instanton number $d$\, where $(Y_I,Z^I)$
has conformal weight $(1+{d\over 2},-{d\over 2})$.

Since the spectral flow operator $F= \d^8(Y)(1-i cu)$ can be used as a
substitute for instanton number, this amplitude can be equivalently
computed with $N$ vertex operators $F(z_r) V(Z(z_r))$ on a disk of
instanton number $d-N$ where $(Y_I,Z^I)$ has conformal weight
$(1+{{d-N}\over 2}, {{N-d}\over 2})$.  But
\eqn{FVdef}
{F(z_r) c(z_r) V(Z(z_r)) = \d^8(Y(z_r)) c(z_r) V(Z(z_r)) =
 c(z_r)\int \de^8 Z\, e^{i Y_I Z^I} V(Z(z_r))}
is the Fourier-transform of $c(z_r)V(Z(z_r))$ defined
in~\eqref{fourv}.  And since $Y_I$ now has conformal weight
$-(N-d-2)/2$, the integration over zero modes involves curves of
degree $(N-d-2)$ in $Y_I$.

So the $N$-point amplitude with $d+1$ negative-helicity gluons and
$(N-d-1)$ positive-helicity gluons can be computed either using the
correlation function of vertex operators $V(Z)$ and degree $d$ curves
in $Z^I$, or equivalently, using the correlation function of vertex
operators $\tilde V(Y)$ and degree $(N-d-2)$ curves in $Y_I$. These
two computations are related by a parity transformation which switches
positive and negative helicities and also switches $Y_I$ with $Z^I$.

It will now be shown that this parity invariance of on-shell
amplitudes can be understood as coming from parity invariance of the
string field theory action.

\subsection{Off-shell parity: the kinetic term}\label{section4.2}

Let us denote $P$ as the parity operator that is responsible for the
interchange of $Z^I$ and $Y_I$. To prove that the kinetic
term~\eqref{kinetic} is invariant under parity transformations, we
need to show that
\eqn{need}
{\langle\Phi|Q|\Phi\rangle =
\langle\Phi|P^\dagger  Q P|\Phi\rangle }
where $P|\Phi\rangle$ is the parity transform of $|\Phi\rangle$.

Since $P$ will be defined to be a unitary transformation, $P^\dagger
= P^{-1}$. So we need to show that the BRST operator commutes with
$P$, i.e. 
\eqn{combrst}
{P^{-1} Q P = Q\,.}
Since $P$ should exchange $Y_I$ with $Z^I$, we shall define
\eqn{ytransf}
{P^{-1} Z^I(z) P = P^{IJ} Y_J(z)\,,\qquad
 P^{-1} Y_I(z) P = P_{IJ} Z^J(z)}
where $P^{IJ}$ and $P_{IJ}$ are constant matrices satisfying $P^{IJ}
P_{JK}= \d^I_K.$ To be a unitary transformation, $P$ must preserve the
OPE's of $Y_I$ with $Z^J$ which implies that the matrix $P^{IJ}$ is
antisymmetric/symmetric when the $IJ$ indices are bosonic/fermionic.

One can check that
\eqn{Ptra}
{P^{-1} J(z) P = - J (z)\,,\qquad
P^{-1} T(z) P = T(z) - \p J (z)}
where $J=Y_I Z^I$ and $T= Y_I \p Z^I $. 
So to commute with 
$$
Q= \int \de z(c (T +T_C) + v J +  c b \p c + c u\p v)\,,
$$
one should define
\eqn{Pdeft}
{\begin{array}{|rcl|rcl|}
\hline
P^{-1} Z^I(z) P &=& P^{IJ} Y_J(z)\,,&
P^{-1} Y_I(z) P &=& P_{IJ} Z^J(z)\,,\\
P^{-1} j^k(z) P &=& j^k(z)\,,& & &\\
\hline
P^{-1} v(z) P &=& -v(z) + \p c(z)\,,&
P^{-1} c(z) P &=& c(z)\,,\\
P^{-1} u(z) P &=& -u(z)\,, &
P^{-1} b(z) P &=& b(z) -\p u(z)\,. \\ \hline
\end{array}}
One can verify that the transformations of~\eqref{Pdeft} preserve the
OPEs of the operators $(Z^I,Y_I,j^k, b,c,u,v)$, so $P$ is a unitary
transformation. By defining the parity transformation as
in~\eqref{Pdeft}, one finds that $P^{-1} Q P =Q$, so the kinetic term
of~\eqref{kinetic} is parity-symmetric.

\subsection{Off-shell parity: the cubic terms}

What about the cubic terms? We will see that the sum of the two cubic
terms is invariant, but the $d=0$ and the $d=1$ terms
in~\eqref{actfour} get interchanged. To see this, first note that the
star product of two string fields, $|\Phi_1(w)\rangle *
|\Phi_2(w)\rangle$, depends in a simple manner on the conformal weight
of $w$. If one twists the conformal weight of $w$ by changing its
background charge, one finds that~\cite{sen}
\eqn{twistc}
{|\Phi_1(w)\rangle * |\Phi_2(w)\rangle \to
  e^{in\s(\pi/2)}|\Phi_1(w)\rangle * |\Phi_2(w)\rangle}
where $w=e^{i\s}$ and $n$ is the shift in the background charge.
Equation~\eqref{twistc} is easily derived from the fact that all
curvature in the cubic vertex is concentrated at the midpoint, so the
exponential of the term $n\int \de^2 z \,\sigma(z) R(z)$\, in the
worldsheet action only contributes at the midpoint.

Under the parity transformation of~\eqref{Pdeft}, the $d=0$ cubic term
$\langle\Phi|(|\Phi\rangle *|\Phi\rangle)$ transforms into
$\langle\Phi|P^{-1} (P|\Phi\rangle * P|\Phi\rangle)$.  So if
$P|\Phi\rangle * P|\Phi\rangle$ were equal to $P(|\Phi\rangle *
|\Phi\rangle)$, the $d=0$ cubic term would be invariant. However,
since $P$ does not commute with the stress tensor $T$, it changes the
conformal weights of the variables and modifies their star product.
Defining
\eqn{stresst}
{T(z) = \{Q, b(z)\} = Y_I \p Z^I + T_C + b \p c +
\p(bc) + u \p v\,,}
one finds that 
\begin{eqnarray}
P^{-1} T(z) P &=& \{Q, P^{-1} b(z) P\} =
\{Q, P^{-1} (b(z) -\p u(z)) P\}
\nonumber\\
&=& T(z) - \p (Y_I Z^I - \p(cu)) = T(z) - \p^2 \s
\label{stressc}
\end{eqnarray}
where $\p\s =Y_I Z^I - \p(cu)$. 

\pagebreak[3]

So the background charge is shifted, which means that 
\eqn{cubg}{
P|\Phi\rangle * P|\Phi\rangle=
P( e^{i\s(\pi/2)}|\Phi\rangle * |\Phi\rangle) =
P\left( F\left(\frac{\pi}{2}\right)|\Phi\rangle * |\Phi\rangle\right) .}
Therefore, the parity transform of 
$\langle\Phi|(|\Phi\rangle *|\Phi\rangle)$ is
$$
\langle\Phi| P^{-1} P\left( F\left(\frac{\pi}{2}\right)|\Phi\rangle *
|\Phi\rangle\right) = \langle\Phi|F\left(\frac{\pi}{2}\right)(|\Phi\rangle
*|\Phi\rangle)\,,
$$
which is the $d=1$ cubic term.

Similarly, the $d=1$ cubic vertex $\langle\Phi|F(\pi/2)(|\Phi\rangle
*|\Phi\rangle)$ transforms into
$$
\langle\Phi|P^{-1}F\left(\frac{\pi}{2}\right)
(P|\Phi\rangle * P|\Phi\rangle)
$$
under a parity transformation. Using~\eqref{cubg}, this is
equal to 
$$
\langle\Phi|P^{-1}F\left(\frac{\pi}{2}\right) P 
\left( F\left(\frac{\pi}{2}\right)|\Phi\rangle * |\Phi\rangle\right).
$$
But one can easily check from the transformation of~\eqref{Pdeft}
that 
$$
P^{-1} F\left(\frac{\pi}{2}\right) P = F^{-1}\left(\frac{\pi}{2}\right)
$$
where $F^{-1}(z) = e^{-i\s(z)}$. Since $F^{-1}(\pi/2) F(\pi/2)=1$, one
finds that $\langle\Phi|F(\pi/2)(|\Phi\rangle *|\Phi\rangle)$
transforms into $\langle\Phi|(|\Phi\rangle *|\Phi\rangle)$ under
parity, which is the $d=0$ cubic vertex.

This is exactly what we want: the two cubic terms in~\eqref{actfour}
are interchanged.  It agrees with the fact that the $d=0$ term couples
the $(++-)$ helicities while the $d=1$ term couples the $(--+)$
helicities, and these two cases are $P$ images of each other.

\section{Conclusions and outlook}\label{outlook}

We have described the string field theory version of the twistorial
open string. The action has a quadratic term, based on the BRST
operator, and two cubic terms. One of these cubic terms contains the
spectral flow operator that is able to increase the degree of the
curve represented by the worldsheet, and the usual string field theory
calculation of scattering amplitudes seems to reproduce $\N=4$
super-Yang-Mills amplitudes transformed into twistor space.

Moreover, our string field theory sheds some new light on the origin
of the parity symmetry that seems non-trivial in the twistor
variables. It would be interesting to see whether our explanation of
parity symmetry can be related to the upcoming papers
of~\cite{wittenp} and~\cite{agavafa}. We would like to list several
other interesting open problems:

\begin{itemize}

\item\emph{Field theory prescriptions:} a twistor-related field theory
  prescription, based purely on degree-one curves connected by a
  propagator, has recently been shown to reproduce Yang-Mills tree
  amplitudes~\cite{svrcek}. We view these results as intriguing, but
  do not yet understand how to relate this prescription to our field
  theory action. Since Yang-Mills states cannot be taken off-shell in
  our formalism, it is unclear how to reproduce the propagators
  of~\cite{svrcek}. A similar difficulty arises in trying to relate
  our action to the spinor helicity methods developed by Chalmers and
  Siegel in~\cite{chalmers}.

\item \emph{Supersymmetric actions:} our field theory action provides
  a manifestly ${\cal N}=4$ supersymmetric method for computing ${\cal
    N}=4$ super-Yang-Mills amplitudes, which has been a longstanding
  open problem using standard superspace approaches. Admittedly, our
  solution of this problem is highly non-conventional since there is
  no natural way to take the theory off-shell and since the
  super-Yang-Mills fields are necessarily coupled to conformal
  supergravity fields.  Nevertheless, our string field theory action
  may give some useful clues for constructing conventional superspace
  actions for ${\cal N}=4$ super-Yang-Mills, perhaps by including
  couplings to conformal supergravity.

\item \emph{Loop diagrams:} although we have only investigated tree
  amplitudes, one can in principle use our string field theory action
  to compute loop amplitudes. There are several new features which are
  expected to arise such as anomalies and closed string poles.
  Hopefully, these new features will help to explain the mysterious
  $c=28$ current algebra and the role of conformal supergravity in the
  open string sector. The question of infrared divergences must also
  be addressed, which is nontrivial in conformal field theories since
  they are coupled at all distance scales.

\item \emph{Off-shell and nonperturbative physics:} it remains to be
  seen whether the prescriptions based on twistor variables can be
  generalized to physics that is off-shell in the usual Minkowski
  space and whether the twistor space ``knows'' about non-perturbative
  physics, for example the S-duality.

\end{itemize}

\vspace{-3mm}

\acknowledgments

\vspace{-3mm}

{\small
We are grateful to Michal Fabinger, Sergei Gukov, Andrew Neitzke,
Warren Siegel, Andrew Strominger, Cumrun Vafa, Anastasia Volovich, and
Edward Witten for useful discussions.  N.B. would like to thank CNPq
grant 300256/94-9, Pronex 66.2002/1998-9, and Fapesp grant 99/12763-0
for partial financial support, and the Institute for Advanced Study
and Harvard University for their hospitality.  The work of L.M. was
supported in part by Harvard DOE grant DE-FG01-91ER40654 and the
Harvard Society of Fellows.
}

\end{document}